          \documentstyle[aps,twocolumn,epsfig,prl]{revtex}

\begin{document}
\draft
         \twocolumn[\hsize\textwidth\columnwidth\hsize\csname
         @twocolumnfalse\endcsname

\title{Discrete saturation thickness and anomalous potential height of native
         ultrathin aluminum oxide tunnel barriers}
\author{K.~Gloos, P.~J.~Koppinen, and J.~P.~Pekola}
\address{Department of Physics, University of Jyv\"askyl\"a, PO Box 35 (YFL), 
                  FIN-40351 Jyv\"askyl\"a, Finland}
\date{\today}
\maketitle

\begin{abstract}
We have investigated planar metal - insulator - metal tunnel junctions with 
aluminum oxide as dielectricum.
These oxide barriers were grown on an aluminum electrode in pure oxygen
at room temperature till saturation.
We observed discrete barrier widths separated by $\Delta s \approx 0.38\,$nm, 
corresponding to the addition of one layer of oxygen atoms.
The minimum thickness of $s_0 \approx 0.54\,$nm is due to a double 
layer of oxygen. 
We found a strong and systematic dependence of the barrier height $\Phi_0$ on 
the thickness $s$ like $\Phi_0 \approx 2.5\,{\rm{eV}} / s^2({\rm{nm}})$, which 
nearly coincides with  the kinetic electron energy $E = h^2/2ms^2$ for which 
the deBroglie wavelength matches the width of the barrier.
\end{abstract}


        ]  
        \narrowtext

Aluminum oxide (AlO$_x$) is one of the standard construction materials for
tunneling barriers because it is easily and reliably being formed. 
It is important for applications like single-electron transistors \cite{SET} and 
Coulomb-blockade thermometers \cite{Pekola94}. 
Aluminum oxide is also a promising candidate to replace SiO$_2$ based gate 
dielectrics in miniaturized electronic circuits, see for example 
Ref.~\cite{Kundu01}.
Although many experiments take advantage of AlO$_x$ tunnel barriers, 
few reports focus on their intrinsic properties.
The conventional point of view on these properties summarizes as follows: 
$i)$ Almost any conductance per area can be achieved by simply choosing the 
       right oxidation pressure and time. 
$ii)$ The typical height of the tunnel barrier is about 2\,eV. But experimental
      data vary from below 0.1\,eV up to 8.6\,eV \cite{Gundlach71,Barner89,Lau81}. 
$iii)$ The dielectric constant $\epsilon$ of the aluminum oxide barrier is 
    smaller than that of bulk Al$_2$O$_3$, which is around 4.5 - 8.9 at around 
    295\,K \cite{handbook83}. Like the height, $\epsilon$ does not seem to be a
   well-defined property of AlO$_x$.

In view of the importance of these tunnel barriers we have investigated
them in more detail.
We show that the height of ultrathin aluminum oxide barriers is {\em not at all} 
related to the band gap of bulk Al$_2$O$_3$. Instead, it seems to be a purely 
geometrical effect set by the lowest kinetic electron energy for which the 
deBroglie wavelength is smaller than the width $s$ of the barrier. 
This interpretation is supported by resolving, for the first time, the 
layer-by-layer thickness of the saturated oxide barrier. 

The most simple picture of a tunnel junction assumes an $s$ wide rectangular
barrier of height $\Phi_0$ across which electrons tunnel. 
In Wentzel-Kramers-Brillouin (WKB) approximation the conductance per area 
at zero bias and at low temperatures becomes \cite{Simmons63}
\begin{equation}
  g_0 = \frac{e^2 \sqrt{2m\Phi_0}}{h^2 s} \, 
             \exp{\left( \frac{2s}{\hbar} \, \sqrt{2m\Phi_0} \right)} .
\label{zero-bias}
\end{equation}
The first order correction increases the conductance quadratically with bias 
voltage $V$ like \cite{Simmons63}
\begin{equation}
  g(V) = g_0\,(1+V^2/V_0^2)
\label{slope}
\end{equation}
with  $V_0^2 = (4 \hbar^2 / e^2 m) \Phi_0 / s^2$. 
Equations \ref{zero-bias} and \ref{slope} can be used to estimate $s$ and
$\Phi_0$  of the junction.
In general, image forces act on the electrons as they move through the 
barrier from one metal electrode to the other. 
They distort the potential  distribution of real  junctions, as described by 
the standard Simmons model  \cite{Simmons63}. 
The barrier height and its width determine then, together with the dielectric 
constant $\epsilon$, the zero-bias conductance $g_0$ and the slope 
parameter $V_0$.
Zero-bias anomalies (ZBA) omitted so far are discussed below.

\begin{figure}[t]
    \includegraphics[width=6.5cm]{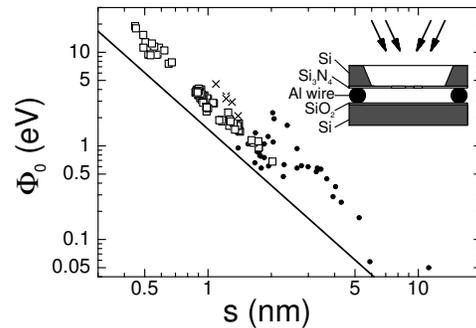}
\caption{Barrier height $\Phi_0$ versus width $s$ of aluminum oxide tunnelling
barriers. The solid line represents Eq.~\ref{deBroglie}. 
Solid circles are the data of Ref.~[5] for sputter-deposited  Al$_2$O$_3$ barriers. 
Crosses are averaged data from Ref.~[6] for thermally or plasma-discharge 
oxidized Al.
The inset shows our experimental setup to evaporate the metallic electrodes
through a thin Si$_3$Ni$_4$ mask onto an oxidized Si substrate (not to scale).
Arrows mark the two different evaporation angles before and after oxidation.}
\label{phi-of-s}
\end{figure}

We fabricated 99 samples on an oxidized Si substrate, using the standard 
two-angle shadow evaporation technique to deposit the metal electrodes
in an ultra-high vacuum (UHV) chamber at a  rate of 0.3\,nm/s.
The samples had  $N = 12$ nominally identical junctions in series, 
each of  $A = 2 \times 2\,\mu{\rm{m}}^2$ cross-section. Electrodes were 
evaporated  through a $300\,$nm thick Si$_3$N$_4$ mask as shown in the
inset of Fig.~\ref{phi-of-s}. 
A $25\,\mu$m diameter Al wire separated the mask from the substrate. 
By using a metal wire as a spacer we avoid organic material near the  sample
which could otherwise contaminate the oxide layer or the electrodes.
The first Al layer, either pure Al or its alloys with Cu, Au, or Nb, was 
exposed to pure oxygen at ambient temperature ($\sim 295\,$K) to form 
a thin insulating AlO$_x$ barrier. 
About $5\cdot 10^4\,$Pa oxygen pressure was applied for 30 minutes in the 
loading chamber of the UHV system.  Then the second electrode 
(Al, Cu, Nb, Au, or a mixture between two of  those metals) was evaporated.
The thickness of  the electrodes was about 90\,nm.
We have found no systematic dependence of the final tunnel conductance
per area on oxygen pressure, oxidation time, evaporation rate, junction area, 
or thickness of the metal electrodes when the relevant parameters were varied 
by a  factor of two. In this respect the experiments are quite reproducible 
and represent the saturation thickness of the barrier at 295\,K in
pure oxygen.

Both the size of the area and the number of junctions are a compromise:
The area is large enough to be reliably measured, but sufficiently small
to show a resolvable Coulomb blockade signal at 4.2\,K when the sample
is in the liquid helium of a transport dewar. 
We have chosen this condition to make a reliable comparison
and to minimize the turn-around time. 
The number of junctions $N$ still allows sufficiently 
large voltages to be applied across each  junction using conventional 
electronic equipment. 
But simultaneously it reduces the risk of external high voltage spikes 
destroying the junctions. 

From the sheet resistance of the shorted junctions - obtained either by  
omitting the oxidation process or when the tunnel barriers have been 
short-circuited by a large bias voltage - we  estimate a residual electrical 
resistivity of the evaporated metals of $4 - 10\,\mu\Omega$cm at 4.2\,K and a 
resistance ratio between 295\,K and 4.2\,K of around 1.5, indicating a good 
quality of the electrodes, both for the pure metals and the alloys. 

The differential conductance $dI/dV$ was recorded by applying a small 
low-frequency ($\sim 100\,$Hz) ac voltage $dV$ superposed on the bias 
voltage $V$ to the junction and detecting the current  modulation $dI$ 
with the help of a lock-in amplifier. 
Results are always presented as the voltage drop per junction and the 
conductance per area $A$ of the junctions $g = (1/A)dI/dV$.

\begin{figure}[t]
     \includegraphics[width=6.5cm]{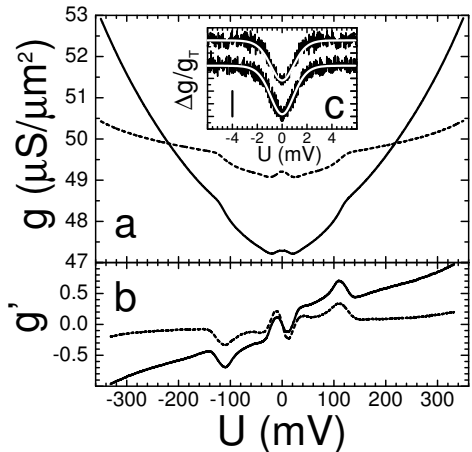}
\caption{a) Differential conductance $g = (1/A)dI/dV$ vs.~voltage drop $V$ for two 
typical tunnel junctions at $T=4.2\,$K. 
Solid line Al-oxide-Cu, dashed line Al-oxide-Au (not symmetrized original data).
b) Second derivative $g' = dg/dU$ of the two spectra in arbitrary units. 
c) Coulomb-blockade anomaly of the two tunnel junctions. The
change $\Delta g$ of conductance due to Coulomb blockade has been normalized 
to the background conductance $g_T$. Solid white lines are theoretical 
predictions. The vertical bar indicates a $2\times 10^{-4}$ variation.
From the size of the anomalies we estimate junction capacitances of about $0.22\,$pF
and $0.18\,$pF and dielectric constants of $\epsilon \approx  5.6$ and 3.1 for 
the Cu (top) and the Au (bottom) sample, respectively.}
\label{spectra}
\end{figure}

Figure \ref{spectra} shows typical $dI(V)/dV$ spectra of an Al-oxide-Cu 
and an Al-oxide-Au tunnel contact at $T=4.2\,$K. Both have the same 
characteristic zero-bias anomalies.
The essential difference between them is the strength of the $V^2$-dependence 
at larger voltages. It is basically this difference which indicates the thinner 
(Au contact) and the thicker (Cu contact) barrier. 

Since most of our junctions have normal electrodes (at $T=4.2\,$K), we rely 
on several observations to ensure quantum tunnelling:
$i)$ The typically $\sim 15\%$ reduction of the conductance on cooling 
from 295\,K to 4.2\,K \cite{Gloos00}, which has been proposed recently as a 
good tunnelling indicator \cite{Jonsson00}.
$ii)$ The $V^2-$dependence of the conductance.
$iii)$ Several zero-bias anomalies (ZBA) appear at lower voltages. They can be
attributed to inelastic electron-phonon scattering in the barrier \cite{Lau81}. 
We believe that the well-pronounced characteristic 110\,meV anomaly, which we 
have regularly observed at thin barriers with $s \le 1.0\,$nm, indicates the 
high quality of those tunnel junctions.
$iv)$ Several Al-oxide-Nb junctions, prepared at the same conditions as the  
other samples, showed  the expected superconducting anomalies.
$v)$ The known shape of the Coulomb blockade spectrum, see below. 

Partial Coulomb blockade at $T = 4.2\,$K can be resolved as long as the 
background zero-bias anomaly has a conductance maximum at $V = 0$. 
Figure \ref{spectra} c) shows that the spectra normalized to the smoothed
background fit well the expected shape due to Coulomb blockade. 
This also indicates that the $N=12$ junctions are rather identical.
In both cases the width of the anomaly \cite{Pekola94}  $V_{1/2} = 
5.44\,k_{\rm{B}} T / e\,$ derived from the fit curve corresponds to a nominal 
temperature of 4.4(3)\,K, in good agreement with the experimental 4.2\,K.
Because of the large contact area the Coulomb blockade anomaly is quite small. 
Nevertheless, its relative size can be used to derive the capacitance of the 
junctions $C$ \cite{Pekola94}.
From that we estimate an  average dielectric constant of $\epsilon \approx 4$ 
for the parallel-plate geometry with $C = \epsilon \epsilon_0 A /s$. 

Surprisingly, most of the tunnel spectra are quite symmetric
like that in Fig.~\ref{spectra}, even when the electrodes are of different 
metals, implying that the tunnel barriers are likewise symmetric. 
Therefore we did not apply the generalized Simmons model for junctions 
with an asymmetric barrier \cite{Brinkman70}, but always symmetrized the 
spectra to obtain the average barrier heights.
Plotting the differential conductance versus $V^2$ reveals a linear 
relationship above the 110\,meV anomaly from which $V_0$ can be extracted
(the Al-oxide-Au samples had a very small $V^2$ contribution, making the
analysis more difficult).
Fitting the spectra above the ZBA using Simmons model and the average 
$\epsilon = 4$  allows us to determine $s$ and $\Phi_0$. 
The final result depends only weakly on the absolute value of $\epsilon$
if this is varied by $\pm 50\%$, and the complete analysis indicates a 
temperature dependence  of $\epsilon$ and/or $\Phi_0$ \cite{Gloos02}.

\begin{figure}[t]
    \includegraphics[width=6cm]{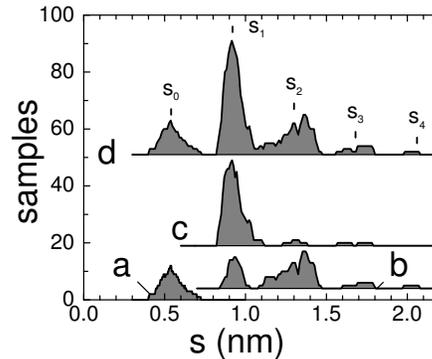}
\caption{Histogram of the number of samples falling inside a 0.1\,nm wide 
interval around a barrier thickness $s$. A total of 99 samples have been
measured. Trace a represents the Al-oxide-Au samples, trace b belongs 
to (M, Al)-oxide-M and (M, Al)-oxide-(Al, M) samples (M = Al, Cu, Au, or Nb), 
and trace c to the Al-oxide-M samples. Trace d is the sum of all samples. 
Traces b, c, and d are vertically displaced.}
\label{histogram}
\end{figure}

Figure \ref{phi-of-s} summarizes our main findings. First, the barrier height
depends very strongly and systematically on the thickness $s$ of the barrier 
like $\Phi_0 \propto 1/s^2$. 
These data clearly demonstrate that $\Phi_0$ is neither related with  the 
band gap of  bulk Al$_2$O$_3$ nor to the work function of the different 
metals. 
Second, data points accumulate at certain values of the thickness, following
a regular pattern with a $\Delta s \approx 0.38(5)\,$nm spacing as shown in the 
histogram in Fig.~\ref{histogram}. 
Such a spacing could be expected for a homogeneously grown oxide layer,
built up basically by adding one layer of oxygen atoms to an already existing 
one (the Al atoms are much smaller than the O atoms). 
The smallest observed thickness $s_0 \approx 0.54(5)\,$nm corresponds then
to two oxygen layers as proposed theoretically in Ref.~\cite{Jennison99}. 
This minimum thickness is also consistent with recent observations of stable 
0.6\,nm thick Al$_2$O$_3$ films grown on a Si substrate \cite{Kundu01}
and 0.59\,nm  Al$_2$O$_3$ films on Ta \cite{Chen94}.
The discreteness of the width $s$ with reasonable absolute values strongly
supports our way to derive the width, and therefore also the barrier height.

We  note here that only Al-oxide-Au junctions were of $s_0$ type, all other
samples with pure Al base electrodes, like Al-oxide-Al and  Al-oxide-Cu, were 
mainly of $s_1$ type, as shown in Fig.~\ref{histogram}.
Samples with Al alloys as base electrode, like 
Cu$_{0.5}$Al$_{0.5}$, were mostly of $s_2$ type, but a considerable 
fraction of them was of $s_1$ and $s_3$ type. 
The barrier of samples with a pure Al base electrode seems to be quite
homogenous, while those of Al alloy samples could be inhomogeneous, 
resulting in an average thickness between  $s_1$ and  $s_2$, as seen in the
histogram.
Obviously, thicker barriers are more easily being formed on Al alloys with their
degraded (irregular) surface structure than on pure Al. 
On the other hand, the preference of the Au samples to form very thin barriers
could result from the fact that Au is difficult to oxidize unlike Al or Cu.
After oxidation, the top-most layer is probably chemisorbed oxygen.
When evaporating Al or Cu as the second electrode, these oxygen 
atoms form  bonds with the metals, but  desorb  easily  when Au is 
evaporated.

Especially the thin oxide layers with $s_0$ and $s_1$ have a well-pronounced
110\,meV zero-bias anomaly as shown in Fig.~\ref{spectra}. This anomaly 
characterizes the surface (or longitudinal) phonons of crystalline Al$_2$O$_3$
(sapphire), see for example Ref.~\cite{Popova00}. 
Thus in these samples the oxygen ions are fully bound in a regular crystal 
lattice and not only chemisorbed. 
Chemisorption may be the case for the thicker barriers. For them the 110\,meV 
anomaly is strongly suppressed and an anomaly at around 50\,meV appears.
A second reason for this suppression could be inhomogeneities in the barrier.
This is consistent with observing noisier spectra at those junctions, 
indicating  more defects.

Several other experiments either contradict or support our findings. 
The supporting ones use a sample structure similar to ours, namely 
metal-oxide-metal tunnel juctions:
Lau and Coleman \cite{Lau81} investigated thermally and plasma-discharge 
grown aluminum oxide barriers while
Barner and Ruggiero \cite{Barner89} sputter-deposited Al$_2$O$_3$ tunnel 
barriers. In both cases the absolute values as well as the thickness dependence 
of the barrier height $\Phi_0(s)$ agree well with our data.
Especially the latter report extends our data set to significantly larger $s$.
The contradicting experiments use different sample setups, 
metal-oxide-semiconductors juctions:
Ballistic electron emission spectroscopy on AlO$_x$ found 3.90(3)\,eV high 
barriers for 8\,nm oxide thickness \cite{Ludeke00}, compared to our 
(extrapolated) 40\,meV. 
Internal photoemission gave $\Phi_0 = 3.25(8)\,$eV for $5 - 15\,$nm wide 
barriers  \cite{Afanasev01}. 
According to Ref.~\cite{Ludeke00} there is almost no difference of the 
barrier height when  AlO$_x$ is replaced by  SiO$_2$. But according to 
Ref.~\cite{Afanasev01} the numbers are  different, namely 4.25(5)\,eV
and 3.25(8)\,eV for SiO$_2$ and  AlO$_x$, respectively.

The different results for the different kind of setups could indicate that 
boundary conditions matter. And this could lead to a tentative
explanation for the well-defined 'cut-off energy' $\Phi_0(s)$ of our samples.
The experimental $\Phi_0(s)$  in Fig.~\ref{phi-of-s} marks roughly the 
kinetic energy
\begin{equation}
  E = \frac{h^2}{2ms^2} \approx  \frac{ 1.5\,{\rm{eV}} }  { s^2 ({\rm{nm}}) }  
\label{deBroglie}
\end{equation}
at which the deBroglie wavelength of the electrons becomes smaller than 
the width $s$ of the barriers. The cut-off would then be simply induced by an
electronic resonance phenomenon inside the barrier.

\begin{figure}[t]
    \includegraphics[width=6cm]{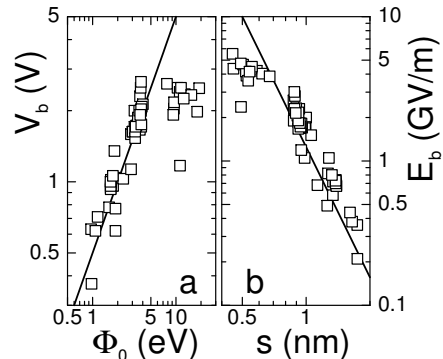}
\caption{a) Breakdown voltage $V_b$ versus barrier height $\Phi_0$ and
b)  breakdown field $E_b$ versus barrier thickness $s$. The solid lines
are $V_b = 0.5\,\Phi_0/e$ and $E_b = 1.25\,({\rm{GV/m}}) / s^3({\rm{nm}})$, 
respectively.}
\label{breakdwn}
\end{figure}

We have  tried to verify independently the potential height by measuring
the current-voltage characteristics to higher voltages. 
Figure \ref{breakdwn} shows the voltage $V_b$ at which the tunnel current
raises dramatically, exceeding $\sim 1\,$mA. 
The junctions are then usually destroyed, resulting in either  a short or an 
open circuit.
However, this breakdown voltage $V_b$ seems to be smaller than the one 
expected for field emission. 
Up to about $\Phi_0 = 4\,$eV the breakdown voltages are closely correlated 
with the potential height, thus  supporting the magnitude of $\Phi_0$ derived 
from the spectra at low bias voltages.
The larger $\Phi_0$ above 4\,eV requires additional considerations because of 
the obvious trend of $V_b$ to saturate at around 2.5\,V (the Al-oxide-Au 
samples had slightly, but systematically, larger $V_b$ than the Al-oxide-Cu or
Al-oxide-Al samples). 
Either these large barrier heights are wrong or we face a new effect. 
The latter is probably the case because of the huge electrical breakdown 
fields $E_b = V_b/s$, approaching $\sim 5\,$GV/m at small $s$. Compared to 
that, typical literature data for AlO$_x$ junctions have much lower 
$E_b \approx 0.1 - 1\,$GV/m at most \cite{Bhatia89}.
A breakdown field which decreases with increasing gap between the 
electrodes is a well-known, yet not fully understood, phenomenon in vacuum 
high-voltage insulation \cite{Diamond98}.
The 2.5\,V upper bound of $V_b$ could be readily explained by the damage 
done due to the mechanical stress excerted by the electrical field.
One could also speculate whether the extremely high tunnel current density 
$\sim 10^8\,$A/m$^2$ just below $V_b$ induces a high enough electron density 
inside the barrier that could transform the insulating AlO$_x$ into a 
conducting metallic state.
Whatever the right explanation, the large $E_b$ confirms  the 
excellent quality of our  tunnel barriers with a very small number of 
defects that could act as weak spots.

Finally, the systematic $\Phi_0 (s)$ dependence leads to a preferred 
transparency of the tunnel junctions for each saturation thickness which 
covers roughly two decades in conductance per area.
A wider range of transparencies is difficult to achieve by just varying the
thickness of the barriers. 

We thank N.~Kim for discussions.
This work has been supported by the Academy of Finland under the Finnish
Centre of Excellence Programme 2000-2005 (Project No. 44875, Nuclear and
Condensed Matter Programme at JYFL) and by the European Commission under
the Measurement and Testing activity of the Competitive and Sustainable Growth
Programme (Contract G6RD-CT-1999-00119).





\begin{thebibliography}{999}
\bibitem{SET}
  H.~Grabert and M.~H.~Devoret (eds.), {\em Single Charge Tunneling}
  (Plenum Press, New York, 1992).
\bibitem{Pekola94}
  J.~P.~Pekola, K.~P.~Hirvi, J.~P.~Kauppinen, and M.~A.~Paalanen,
  Phys.~Rev.~Lett.~{\bf 73}, 2903 (1994).
\bibitem{Kundu01}
  M.~Kundu, N.~Miyata, and M.~Ichikawa,
  Appl.~Phys.~Lett.~{\bf 78}, 1517 (2001).
\bibitem{Gundlach71}
  K.~H.~Gundlach and J.~H\"olzl,
  Surface Science {\bf 27}, 125 (1971).
\bibitem{Barner89}
  J.~B.~Barner and S.~T.~Ruggiero, Phys.~Rev.~{\bf B 39}, 2060 (1989).
\bibitem{Lau81}
  J.~C.~Lau and R.~V.~Coleman, Phys.~Rev.~{\bf B 24}, 2985 (1981).
\bibitem{handbook83}
  R.~E.~Bolz and G.~L.~Tuve (eds.), {\em CRC Handbook of Tables for 
  Applied Engineering Science},
  (CRC Press, Boca Raton, Florida, 1983).
\bibitem{Simmons63}
 J.~G.~Simmons, J.~Appl.~Phys.~{\bf 34}, 1793 (1963).
\bibitem{Gloos00}
  K.~Gloos, R.~S.~Poikolainen, and J.~P.~Pekola,
  Appl.~Phys.~Lett. {\bf 77}, 2915 (2000).
\bibitem{Jonsson00}
  B.~J.~J{\"o}nsson-{\AA}kerman, R.~Escudero, C.~Leighton, S.~Kim, 
  I.~K.~Schuller, and D.~A.~Rabson,
  Appl.~Phys.~Lett.~{\bf 77}, 1870 (2000).
\bibitem{Brinkman70} 
   W.~F.~Brinkman, R.~C.~Dynes, and J.~M.~Rowell,
  J.~Appl.~Phys. {\bf 41}, 1915 (1970).
\bibitem{Gloos02}
  K.~Gloos {\em et al.}, to be publ.
\bibitem{Jennison99}
  D.~R.~Jennison, C.~Verdozzi, P.~A.~Schultz, and M.~P.~Sears,
  Phys.~Rev.~{\bf B 59}, R15605 (1999).
\bibitem{Chen94}
  P.~J.~Chen and D.~W.~Goodman, 
  Surf.~Sci.~{\bf 312}, L767 (1994).
\bibitem{Popova00}
  I.~Popova, V.~Zhukov, and J.~T.~Yates, Jr.
  J.~Appl.~Phys.~{\bf 87}, 8143 (2000).
\bibitem{Ludeke00}
  R.~Ludeke, M.~T.~Cuberes, and E.~Cartier,
  Appl.~Phys.~Lett.~{\bf 76}, 2886 (2000).
\bibitem{Afanasev01}
  V.~V.~Afanas'ev, M.~Houssa, A.~Stesmans, and M.~M.~Heyns,
  J.~Appl.~Phys.~{\bf 78}, 3073 (2001).
\bibitem{Bhatia89}
  C.~S.~Bhatia, G.~Guthmiller, and A.~M.~Spool,
  J.~Vac.~Sci.~Technol.~{\bf A 7}, 1298 (1989).
\bibitem{Diamond98}
  W.~T.~Diamond,
  J.~Vac.~Sci.~Technol. {\bf A 16}, 707 (1998).

\end{thebibliography}
\end{document}